\documentclass[12pt,draftcls,journal,onecolumn]{IEEEtran}	% Review options

\ifCLASSINFOpdf
\else
\fi
\usepackage[cmex10]{amsmath}
\ifCLASSOPTIONcompsoc
  \usepackage[caption=false,font=normalsize,labelfont=sf,textfont=sf]{subfig}
\else
  \usepackage[caption=false,font=footnotesize]{subfig}
\fi
\newcommand{\matr}[1]{\mathbf{#1}}
\usepackage[exponent-product = \cdot]{siunitx}
\DeclareMathOperator{\hermit}{H}

\usepackage{pgfplots}
\pgfplotsset{compat=newest}
\newlength\figureheight	% Size for plots
\newlength\figurewidth
\newcommand{\columnplot}{
\setlength\figureheight{0.25\textwidth}
\setlength\figurewidth{0.35\textwidth}
}	%ratio 3/4

\usetikzlibrary{external} 
\tikzexternaldisable

\begin{document}
\title{Array Size Reduction for High-Rank LOS MIMO ULAs}

\author{Tim~H{\"a}lsig~and~Berthold~Lankl% <-this % stops a space
%\thanks{Manuscript received April 19, 2005; revised September 17, 2014.}
%\thanks{The work of T.~H{\"a}lsig was supported in part by the German Research Foundation (DFG) in the framework of priority program SPP~1655 ``Wireless Ultra High Data
%Rate Communication for Mobile Internet Access''.}% <-this % stops a space
%\thanks{This work was supported in part by the German Research Foundation (DFG) in the framework of priority program SPP~1655 ``Wireless Ultra High Data Rate Communication for Mobile Internet Access''.}% <-this % stops a space
\thanks{The authors are with the Institute for Communications Engineering, Universit{\"a}t der Bundeswehr M{\"u}nchen, Germany (e-mail: tim.haelsig@unibw.de; berthold.lankl@unibw.de).}% <-this % stops a space
%\thanks{Color versions of one or more of the figures in this paper are available online at http://ieeexplore.ieee.org.}
%\thanks{Digital Object Identifier 12345.de}
}


%\markboth{IEEE Wireless Communications Letters~Vol.~XX, No.~XX, DATE}%
%{H{\"a}lsig \MakeLowercase{\textit{et al.}}: TITLE}
\markboth{}%
{}



\maketitle

\begin{abstract}
In this paper we propose an extended LOS MIMO channel model, which considers an additional phase shifting term in the transmission path, and which provides the potential to improve channel conditioning significantly. We show that this phase shifting can, for example, be achieved by adding a dielectric material between the transmitting and receiving antennas, where the phase shift is dependent on the distance the waves travel in the medium. Using that distance as a design parameter we demonstrate that the optimal spacing between antenna elements of uniform linear arrays, achieving full spatial multiplexing, can be reduced compared with the well-known spacing criterion from previous investigations.
\end{abstract}

%\begin{IEEEkeywords}
%IEEEtran, journal, \LaTeX, paper, template.
%\end{IEEEkeywords}

% \fi
\IEEEpeerreviewmaketitle

\section{Introduction}
\IEEEPARstart{I}{ncreasing} demand for very high spectral efficiencies has generated a lot of research activity in the field of multiple-input multiple-output~(MIMO) techniques. Aside from the regularly used rich-scattering (or Rayleigh fading) channel assumption, it has also become well known that line-of-sight~(LOS) channel conditions can likewise achieve high spectral efficiencies, when certain geometrical rules are obeyed \cite{Driessen1999,Calabro2003}. However, for many practical systems the required geometric setup can preclude these gains as, e.g., the required spacing between antennas can be tremendous. When the setup differs from the optimal arrangement, the conditioning of the channel may weaken considerably \cite{Bohagen2007}, ultimately leading to a significant decrease in spectral efficiency.
In this letter we present a slightly modified channel model which makes use of a dielectric medium in order to improve channel conditioning, when the geometrical rules cannot be complied with. Consequently, this also leads to new optimal antenna spacing requirements which are discussed in this work. Finding medium shapes that yield well-conditioned channels is highly dependent on the antenna setup and on the constraints that are applied to the medium. We provide one approach to tackle this problem and show some results.

The design of LOS MIMO systems that can benefit from spatial multiplexing to achieve very high spectral efficiencies is, among others, discussed in \cite{Bohagen2007} and \cite{Wang2014}. The number of parallel spatial streams that these systems can support is determined by two parameters, the antenna array arrangement, i.e., size and orientation of the arrays at transmitter and receiver, and the wavelength-transmission range product. The general trade-off between the two parameters states that when additional spatial streams need to be supported, either array size has to be increased or, wavelength or transmission range have to be decreased. The work in \cite{Walkenhorst2007} investigates how far-field phase responses of antennas might be beneficial for maximizing capacity in LOS environments. It was shown that this effect can enhance capacity, by improving channel matrix conditioning, depending on the slope of the phase response. Antenna arrays that are less prone to suboptimal geometrical arrangement were examined in \cite{Torkildson2009}. The results reveal that nonuniformly spaced arrays can prove advantageous in these situations and lead to a more stable channel matrix. Sparse MIMO channels were investigated in \cite{Raghavan2011}, unveiling that in the low
SNR regime sub-$\lambda/2$ spacing and beamforming are helpful in terms of capacity, whereas for high SNRs spatial multiplexing achieves the highest gains. An approach exploiting these results is presented in \cite{Brady2013}, where a dielectric lens is used to perform part of the beamforming task. The use of lenses at millimeter-wave frequencies in general has become quite common, e.g., to build highly directional antennas. An implementation of a dielectric lens may, for example, be found in \cite{Imbert2015}.

\section{System Model}
Consider the equivalent baseband representation of a MIMO system to be given by
\begin{equation}
\matr{y} = \matr{H}\matr{x} + \matr{w} 
\end{equation}
where $\matr{x}$, $\matr{y}$ are transmit and receive vector, and $\matr{w}$ is the additive noise vector with complex Gaussian distribution. In this work we will assume $\matr{H} = \matr{H}_\text{LOS}$ meaning that the channel matrix is solely determined by a LOS component which does not vary with time. It has been shown \cite{Bohagen2007} that in such a case, the normalized channel entries are determined entirely by the phase shift that the waves experience while traveling from transmitter to receiver, i.e.,
\begin{equation}
\left(\matr{H}_\text{LOS}\right)_{m,n} = \exp\left(-j\frac{2\pi}{\lambda}r_{mn}\right) \label{eq:Hlos}
\end{equation}
with $\left(\cdot\right)_{m,n}$ denoting the $m$th row and $n$th column matrix entry, $\lambda$ denoting the propagating wavelength and $r_{mn}$ denoting the distance from $n$th transmit to $m$th receive antenna.

In this work we consider the LOS channel matrix to be composed of two parts as
\begin{equation}
\matr{H}_\text{LOS} = \matr{H}_\text{FS}\circ\matr{H}_\text{PS}
\end{equation}
where $\circ$ is the Hadamard product. The two parts contribute as follows, $\matr{H}_\text{FS}$ incorporates the phase shift that the waves gain from free-space propagation, i.e., is equal to \eqref{eq:Hlos} when $\lambda$ is set to the free-space wavelength. In $\matr{H}_\text{PS}$ phase shifts that occur besides the free-space propagation part can be captured, e.g., the waves pass through different dielectric media between transmitter and receiver.

\begin{figure}[!t]
\centering
\resizebox{0.425\textwidth}{!}{\begin{tikzpicture}[>=latex,font=\normalsize]

\begin{scope}[shift={(0,-0.25)},rotate=15]

\draw[line width=1pt] (1.5,0) -| (1.5,1.25);
\draw[line width=1pt] (1.15,1.25) -- (1.5,0.75);
\draw[line width=1pt] (1.85,1.25) -- (1.5,0.75);

\begin{scope}[shift={(0,1.75)}]

\draw[line width=1pt] (1.5,0) -| (1.5,1.25);
\draw[line width=1pt] (1.15,1.25) -- (1.5,0.75);
\draw[line width=1pt] (1.85,1.25) -- (1.5,0.75);

\end{scope}

\begin{scope}[shift={(0,3.5)}]

\draw[line width=1pt] (1.5,0) -| (1.5,1.25);
\draw[line width=1pt] (1.15,1.25) -- (1.5,0.75);
\draw[line width=1pt] (1.85,1.25) -- (1.5,0.75);

\end{scope}

\begin{scope}[shift={(-0.75,3)}]
\draw[<->,>=latex] (1.5,0) -- (1.5,1.75) node [pos=0.5, fill=white,xshift=0mm,yshift=0mm] {$d_t$};
\end{scope}

\end{scope}

\draw[line width=1pt,dashed] (1.5,0.8) arc (90:105:1);
\draw[line width=1pt,dashed] (1.5,0.8) -- (1.5,0.2);
\node[] at (1.75,0.7)  {$\theta_{t}$};

\begin{scope}[shift={(8.5,0)}]

\draw[line width=1pt] (1.5,0) -| (1.5,1.25);
\draw[line width=1pt] (1.15,1.25) -- (1.5,0.75);
\draw[line width=1pt] (1.85,1.25) -- (1.5,0.75);

\begin{scope}[shift={(0,1.75)}]

\draw[line width=1pt] (1.5,0) -| (1.5,1.25);
\draw[line width=1pt] (1.15,1.25) -- (1.5,0.75);
\draw[line width=1pt] (1.85,1.25) -- (1.5,0.75);

\end{scope}

\begin{scope}[shift={(0,3.5)}]

\draw[line width=1pt] (1.5,0) -| (1.5,1.25);
\draw[line width=1pt] (1.15,1.25) -- (1.5,0.75);
\draw[line width=1pt] (1.85,1.25) -- (1.5,0.75);

\end{scope}

\end{scope}

%\begin{scope}[shift={(9.25,1.25)}]
%\draw[<->,>=latex] (1.5,0) -- (1.5,3.5) node [pos=0.59, fill=white,xshift=0mm,yshift=0mm] {$D_r$};
%\end{scope}

\begin{scope}[shift={(0,3.5)}]
\draw[<->,>=latex] (1.25,1) -- (9.375,1) node [pos=0.5, fill=white,xshift=0mm,yshift=0mm] {$r_{11}$};
\end{scope}

\draw[<->,>=latex] (1.25,4.4) -- (9.375,1) node [pos=0.5, fill=white,xshift=0mm,yshift=0mm] {$r_{31}$};

\begin{scope}[shift={(0,-0.75)}]
\draw[<->,>=latex] (2.125,1) -- (9.375,1) node [pos=0.5, fill=white,xshift=0mm,yshift=0mm] {$R$};
\end{scope}

\node[rectangle, minimum width=0.5cm, minimum height=4.25cm,draw=black,fill=black!10] at (8.75,2.75) {};

\node[] at (8.75,4.5) {\color{black}$l_{11}$};
\node[] at (8.75,1.3) {\color{black}$l_{31}$};

\node[align=center] at (8.75,5.25) {Medium, $\epsilon_r$};

\begin{scope}[shift={(0,1.75)}]
\draw[<->,>=latex] (8.5,1) -- (9.0,1) node [pos=0.5,above, fill=none,xshift=0mm,yshift=0mm] {$t$};
\end{scope}

\end{tikzpicture}}
\caption{Generic LOS MIMO setup with distance parameters and medium introducing an additional phase shift in the channel.}
\label{fig:setup}
\end{figure}
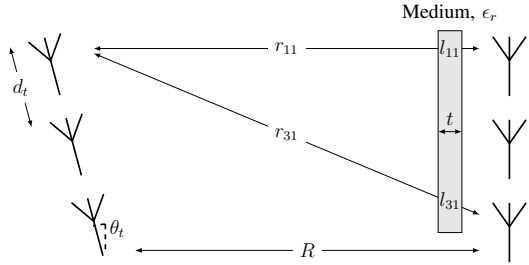

\section{Phase Shift Matrix}
If the entries in the phase shifting matrix $\matr{H}_\text{PS}$ could be designed individually, then a well-conditioned channel matrix $\matr{H}_\text{LOS}$ could always be achieved, independent of the antenna arrangement. One physical effect that achieves a phase shift and can be expressed through the matrix $\matr{H}_\text{PS}$ is the phase response of the antennas due to different angles of arrival, as has been mentioned in \cite{Walkenhorst2007}.

In this work, we will consider a different effect which depends on the fact that the wavelength in different media is different with 
\begin{equation}
\lambda \approx \frac{\lambda_0}{\sqrt{\epsilon_r}}
\end{equation}
where $\lambda_0$ is the wavelength in free-space, $\epsilon_r$ is the relative permittivity of the medium, and the approximation holds if the relative permeability of the medium is close to one. Depending on the distance the waves travel through the medium, different phase shifts can be achieved. We can write the phase shift matrix as
\begin{equation}
\left(\matr{H}_\text{PS}\right)_{m,n} = \exp\left(-j\frac{2\pi}{\lambda}l_{mn}\right) = \exp\left(-j\frac{2\pi}{\lambda_0}\sqrt{\epsilon_r}\cdot l_{mn}\right)
\end{equation}
with $l_{mn}$ being the distance the waves travel in the medium. For example, consider the setup given in Fig.~\ref{fig:setup} where each wave from antenna 1 will experience a different additional phase shift due to the different distances that it travels in the medium. Both $l_{mn}$ and $\epsilon_r$\footnote{It is also possible to have different $\epsilon_r$ values within the medium, which is not going to be considered here.} are design parameters that can be used to achieve the desired phase relations at the receiver.

For the general LOS setup, where the waves travel some distance through a different medium, we can further specify the entries of the complete channel matrix as
\begin{align}
\left(\matr{H}_\text{LOS}\right)_{m,n} = \exp\left(-j\frac{2\pi}{\lambda_0}(r_{mn}-l_{mn})\right) \cdot\exp\left(-j\frac{2\pi}{\lambda_0}\sqrt{\epsilon_r}\cdot l_{mn}\right)
\end{align}
where $r_{mn}-l_{mn}$ is the distance the waves travel in free-space and $l_{mn}$ is the distance the waves travel in the medium between the corresponding antennas. As described in \cite{Bohagen2007} and references therein, an optimal (orthogonal) channel matrix is achieved if the received vectors from two different transmitting antennas are orthogonal, i.e.,
\begin{align}
\left\langle \matr{h}_k,\matr{h}_l \right\rangle = \sum_{m=1}^{M-1} \exp\left(-j\frac{2\pi}{\lambda_0}(r_{mk}-r_{ml})\right) \cdot\exp\left(-j\frac{2\pi}{\lambda_0}(\sqrt{\epsilon_r}-1)(l_{mk}-l_{ml})\right) = 0 \text{.} \label{eq:innerproduct}
\end{align}

\section{Uniform Linear Arrays for LOS MIMO}
It is known \cite{Bohagen2007} that the optimal spacing for uniform linear arrays (ULAs) supporting maximum spatial multiplexing and hence capacity in LOS MIMO systems is determined by 
\begin{equation}
d_t d_r = \frac{\lambda R}{V\cos\theta_t\cos\theta_r} \label{eq:optULA}
\end{equation}
where $R$ is the desired link distance and where $V=\max(N,M)$, with $N$ being the number of transmit antennas and $M$ being the number of receive antennas. Furthermore, $\theta_t$ and $\theta_r$ represent tilting angles at the transmitter and receiver side, as seen in Fig.~\ref{fig:setup}. Note that this equation can lead to relatively large antenna spacings, if the $\lambda R$ product is large.

When the spacing is chosen smaller than specified in \eqref{eq:optULA}, the term
\begin{equation}
1/\kappa = \frac{\lambda_{\min}\left(\matr{H}_\text{LOS}\matr{H}^{\hermit}_\text{LOS}\right)}{\lambda_{\max}\left(\matr{H}_\text{LOS}\matr{H}^{\hermit}_\text{LOS}\right)}
\end{equation}
will be much smaller than one, with $1/\kappa$ being the inverse of the (squared) condition number. Here, $(\cdot)^{\hermit}$ is the conjugate transpose of a matrix and $\lambda_{\min}(\cdot)$, $\lambda_{\max}(\cdot)$ gives the smallest and largest eigenvalue of a matrix, respectively. Note that in the best case, a system designed according to \eqref{eq:optULA}, minimum and maximum eigenvalue are equal, $1/\kappa=1$, supporting full spatial multiplexing. Thus, the term $1/\kappa$ can be used to evaluate the spatial multiplexing capabilities of LOS MIMO systems.

\subsection{Rectangular Medium} \label{sec:recmedium}
For a setup with a rectangular dielectric medium that is parallel to the ULA, as depicted in Fig.~\ref{fig:setup}, the distances in the medium can be calculated through trigonometry by
\begin{equation}
l_{mn} = t\cdot\frac{r_{mn}}{R} \label{eq:l_rec}
\end{equation}
where $t$ is the thickness of the medium. Then, \eqref{eq:innerproduct} can be reformulated as
\begin{align}
\left\langle \matr{h}_k,\matr{h}_l \right\rangle = \sum_{m=1}^{M-1} \exp\left(-j\frac{2\pi}{\lambda_0}\left(1+(\sqrt{\epsilon_r}-1)\frac{t}{R}\right)\right. \left.\cdot\frac{d_t d_r\cos\theta_t\cos\theta_r}{R}(k-l)m\right) \text{,}
\end{align}
applying the same reasoning and approximations as in \cite{Bohagen2007} and references therein (see also the Appendix). We can then find the optimal spacing, yielding $1/\kappa=1$, for ULAs supplemented by a rectangular dielectric medium in
\begin{equation}
d_t d_r = \frac{\lambda_0 R^2}{V(R+t(\sqrt{\epsilon_r}-1))\cos\theta_t\cos\theta_r} \text{.} \label{eq:newspacing}
\end{equation}
It can be seen that in this case, the medium will only have an impact if the term $t(\sqrt{\epsilon_r}-1)$ is significant with respect to $R$. The benefit of adding a rectangular dielectric will thus be most pronounced for short link ranges. For longer ranges the medium may need to be very thick (depending on $\epsilon_r$) in order to be significant. Although the relation still holds for that case, other effects that occur due to the medium, such as refraction and attenuation, should then also be considered, see the Appendix for some additional comments on that.

\subsection{Optimization for two and three Antenna ULAs} \label{sec:opt23}
Finding the general $l_{mn}$ values, i.e. no specific medium shape assumed, which are subject to some physical constraints is very complicated even for ULAs. We will thus consider that the matrix $\matr{L}$, which includes all lengths that the waves travel in the medium between corresponding antennas, has a Toeplitz structure with
\begin{equation}
\matr{L}=
\begin{bmatrix}
l_{11} & l_{12} & \cdots & l_{1N} \\
l_{12} & l_{11} & \cdots & l_{1N-1} \\
\vdots & \vdots & \ddots & \vdots   \\
l_{1N} & l_{1N-1} & \cdots & l_{11} \\
\end{bmatrix} \text{.}
\label{eq:toeplitz}
\end{equation}
This is a valid assumption if the antennas are uniformly spaced and the arrays are aligned (no rotation), if the shape of the medium is symmetric with respect to the center of each antenna, and if the same shape of medium is used for each antenna\footnote{Assume, e.g., the same dielectric lens in front of each receiving antenna.} (see also the Appendix). When $M>N$ or $N>M$ the structure can also be used by simply omitting the corresponding columns or rows of the Toeplitz matrix. 

\begin{figure*}[!t]
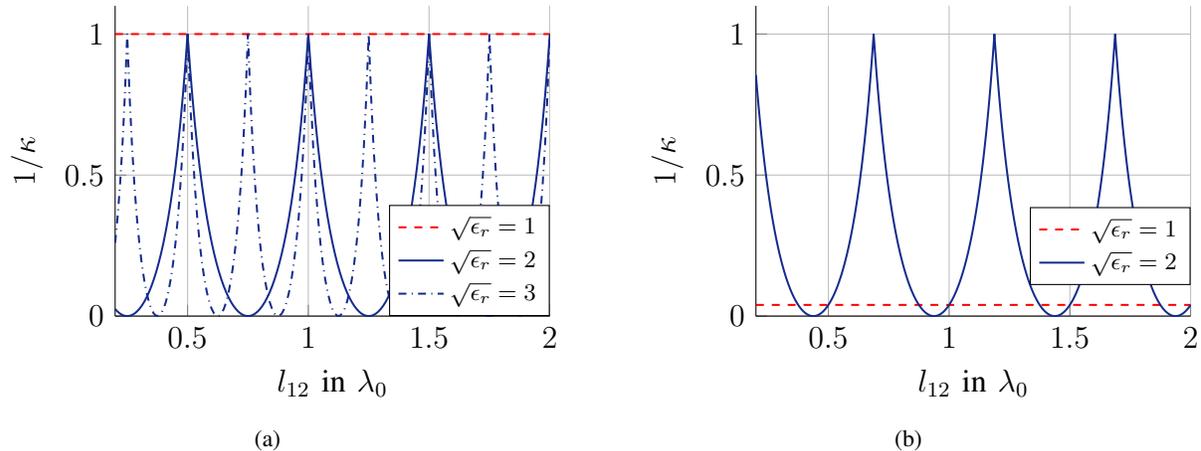

\centering
\subfloat[]{\columnplot\input{figures/ULAopt_standard}\label{fig:ULAopt_standard}}
\hfill
\tikzexternalenable
\tikzsetnextfilename{2antennaOptimization}
\subfloat[]{\columnplot\input{figures/ULAopt_050d}\label{fig:ULAopt_050d}}
\tikzexternaldisable
\caption{Inverse condition number for two antenna LOS MIMO ULAs, $M=N=2$, versus the distance in different dielectric media and for different antenna spacings, $l_{11}=\lambda_0$: \protect\subref{fig:ULAopt_standard}~Optimal spacing of $1.0\cdot d_\text{Opt}$; \protect\subref{fig:ULAopt_050d}~Reduced antenna spacing of $0.5\cdot d_\text{Opt}$.}
\label{fig:ulaopt_2}
\end{figure*}

In Fig.~\ref{fig:ulaopt_2} the inverse condition number for the symmetric two antenna case is shown, to illustrate the reduction in spacing we define $d_t=d_r=d_\text{Opt}$. The first plot shows the case when the optimal spacing is fulfilled, where as expected condition number is always one if $\sqrt{\epsilon_r}=1$ meaning that the medium is free-space. When $\sqrt{\epsilon_r}>1$ the condition number becomes one for cases where the phase shift through the dielectric medium and through free-space again satisfies the orthogonality criterion. The number of points which achieve $1/\kappa=1$ increases with $\sqrt{\epsilon_r}$ as more phase shift is achieved over a given distance. In the second plot of Fig.~\ref{fig:ulaopt_2} the antenna spacing is set to half of the optimum. We can see that the standard case without medium, $\sqrt{\epsilon_r}=1$, has a significantly reduced condition number. However, when a dielectric medium is used the condition number is improved over a wide range of values and $1/\kappa=1$ is still achievable.

\begin{figure}[!t]
\centering
\columnplot
\tikzexternalenable
\tikzsetnextfilename{3antennaOptimization}
\input{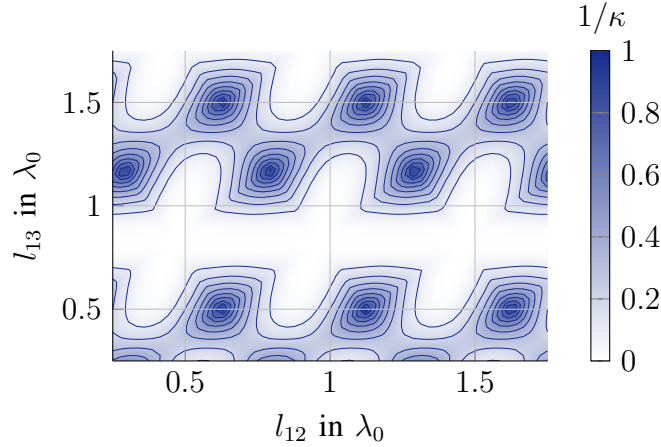}
\tikzexternaldisable
\caption{Inverse condition number for three antenna LOS MIMO ULAs, $M=N=3$, for different distances in a dielectric medium with $\sqrt{\epsilon_r}=2$,  $l_{11}=\lambda_0$, and reduced antenna spacing of $0.5\cdot d_\text{Opt}$.}
\label{fig:ULAopt_050d_3ant}
\end{figure}

Fig.~\ref{fig:ULAopt_050d_3ant} shows the inverse condition number for a three antenna ULA when reduced spacing and a medium with $\sqrt{\epsilon_r}=2$ are used. It can also be seen here that an optimal condition number of one is obtainable, for reference without the medium $1/\kappa \approx 10^{-3}$.
Note that the results are not independent of the parameters used in \eqref{eq:optULA}, but rather they show that for a given spacing reduction the same lengths can be applied. Compared with the results of Sec.~\ref{sec:recmedium}, we find that the numerically determined optimal values for $l_{mn}$ here, will also appear in \eqref{eq:l_rec} when $t$ is chosen optimally.

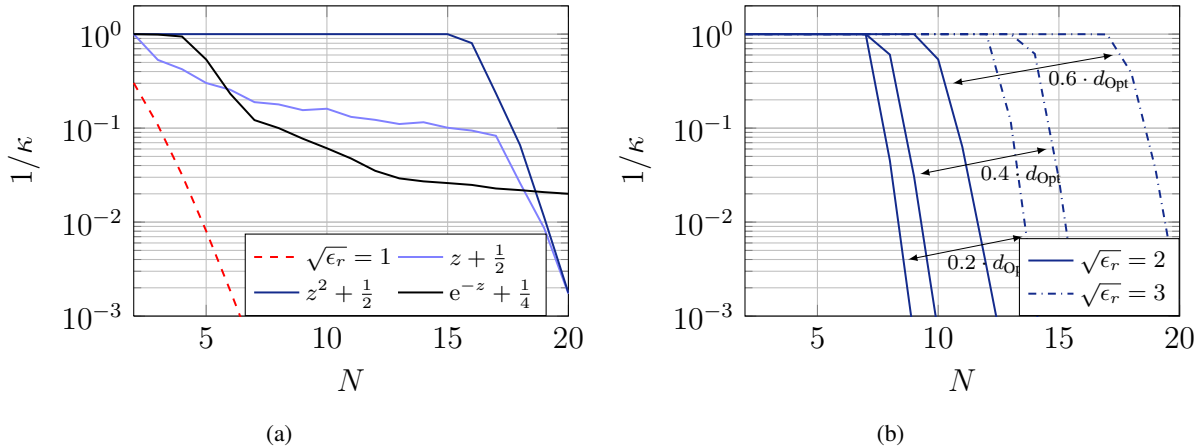
\begin{figure*}[!t]
\centering
\subfloat[]{\columnplot% This file was created by matlab2tikz v0.4.7 running on MATLAB 8.3.
% Copyright (c) 2008--2014, Nico Schlömer <nico.schloemer@gmail.com>
% All rights reserved.
% Minimal pgfplots version: 1.3
% 
% The latest updates can be retrieved from
%   http://www.mathworks.com/matlabcentral/fileexchange/22022-matlab2tikz
% where you can also make suggestions and rate matlab2tikz.
% 
%
% defining custom colors
\definecolor{mycolor1}{rgb}{0.0784313753247261,0.168627455830574,0.549019634723663}%
\begin{tikzpicture}

\begin{axis}[%
width=\figurewidth,
height=\figureheight,
scale only axis,
xmin=2,
xmax=20,
xmajorgrids,
xlabel={$N$},
ymode=log,
ymin=1e-3,
ymax=2,
yminorticks=true,
ymajorgrids,
yminorgrids,
ylabel={$1/\kappa$},
legend columns=2,
legend style={at={(0.95,0.0)},anchor=south east,draw=black,fill=white,legend cell align=left,font=\footnotesize}
]
\addplot [color=red,dashed,line width=0.75pt]
  table[row sep=crcr]{2	0.302227299\\
3	0.107721657\\
4	0.031811777\\
5	0.008045181\\
6	0.001854766\\
7	0.00040401\\
8	8.49e-05\\
9	1.74e-05\\
10	3.51e-06\\
11	6.99e-07\\
12	1.37e-07\\
13	2.68e-08\\
14	5.18e-09\\
15	9.97e-10\\
16	1.91e-10\\
17	3.63e-11\\
18	6.88e-12\\
19	1.3e-12\\
20	2.45e-13\\
};
\addlegendentry{$\sqrt{\epsilon_r}=1$};

\addplot [color=blue!50,solid,line width=0.75pt]
  table[row sep=crcr]{2	0.999991958\\
3	0.531961962\\
4	0.423593445\\
5	0.302102481\\
6	0.255957146\\
7	0.18902264\\
8	0.178801837\\
9	0.155381276\\
10	0.160792252\\
11	0.131716626\\
12	0.122465943\\
13	0.110551967\\
14	0.115505998\\
15	0.100921588\\
16	0.094166528\\
17	0.082846702\\
18	0.025970896\\
19	0.00859336\\
20	0.001732449\\
};
\addlegendentry{$z+\frac{1}{2}$};

\addplot [color=mycolor1,solid,line width=0.75pt]
  table[row sep=crcr]{2	0.999978845\\
3	0.999871649\\
4	0.999973445\\
5	0.9999421\\
6	0.999777686\\
7	0.999298774\\
8	0.999612474\\
9	0.99957629\\
10	0.999696529\\
11	0.999507631\\
12	0.999615217\\
13	0.999579099\\
14	0.999486882\\
15	0.999399135\\
16	0.802590687\\
17	0.234298218\\
18	0.065834044\\
19	0.011206314\\
20	0.001766724\\
};
\addlegendentry{$z^2+\frac{1}{2}$};

\addplot [color=black,solid,line width=0.75pt]
  table[row sep=crcr]{2	0.999828038\\
3	0.989429317\\
4	0.9431069\\
5	0.537480945\\
6	0.231848335\\
7	0.121739871\\
8	0.100152762\\
9	0.077232351\\
10	0.061031575\\
11	0.047530408\\
12	0.035142171\\
13	0.029192247\\
14	0.027132856\\
15	0.026024052\\
16	0.024870626\\
17	0.02278723\\
18	0.021885704\\
19	0.020825408\\
20	0.020034929\\
};
\addlegendentry{$\mathrm{e}^{-z}+\frac{1}{4}$};

\end{axis}
\end{tikzpicture}%
\label{fig:OptFunctions}}
\hspace{-4mm}
%\subfloat[]{\columnplot\input{figures/SqEpsilonr}\label{fig:SqEpsilonr}}
\subfloat[]{\tikzexternalenable\tikzsetnextfilename{SqEpsilonr}\columnplot% This file was created by matlab2tikz v0.4.7 running on MATLAB 8.3.
% Copyright (c) 2008--2014, Nico Schlömer <nico.schloemer@gmail.com>
% All rights reserved.
% Minimal pgfplots version: 1.3
% 
% The latest updates can be retrieved from
%   http://www.mathworks.com/matlabcentral/fileexchange/22022-matlab2tikz
% where you can also make suggestions and rate matlab2tikz.
% 
\definecolor{mycolor1}{rgb}{0.0784313753247261,0.168627455830574,0.549019634723663}%

\begin{tikzpicture}
\tikzset{>=latex}
\begin{axis}[%
width=\figurewidth,
height=\figureheight,
scale only axis,
xmin=2,
xmax=20,
xmajorgrids,
xlabel={$N$},
ymode=log,
ymin=1e-3,
ymax=2,
yminorticks=true,
ymajorgrids,
yminorgrids,
ylabel={$1/\kappa$},
legend style={at={(1.00,0.0)},anchor=south east,draw=black,fill=white,legend cell align=left,font=\footnotesize}
]
%\addplot [color=red,dashed,line width=0.75pt,forget plot]
  %table[row sep=crcr]{2	0.302227299\\
%3	0.107721657\\
%4	0.031811777\\
%5	0.008045181\\
%6	0.001854766\\
%7	0.00040401\\
%8	8.49e-05\\
%9	1.74e-05\\
%10	3.51e-06\\
%11	6.99e-07\\
%12	1.37e-07\\
%13	2.68e-08\\
%14	5.18e-09\\
%15	9.97e-10\\
%16	1.91e-10\\
%17	3.63e-11\\
%18	6.88e-12\\
%19	1.3e-12\\
%20	2.45e-13\\
%};
\addplot [color=mycolor1,solid,line width=0.75pt]
  table[row sep=crcr]{2	0.999774586\\
3	0.999480918\\
4	0.999791801\\
5	0.999312639\\
6	0.999670964\\
7	0.999264248\\
8	0.999151883\\
9	0.999935363\\
10	0.53853204\\
11	0.063269208\\
12	0.003364974\\
13	0.000162416\\
14	8.15e-06\\
15	4.25e-07\\
16	2.26e-08\\
17	1.21e-09\\
18	6.57e-11\\
19	3.56e-12\\
20	1.92e-13\\
};
\addlegendentry{$\sqrt{\epsilon_r}=2$};

\addplot [color=mycolor1,solid,forget plot,line width=0.75pt]
  table[row sep=crcr]{2	0.999781837\\
3	0.999720404\\
4	0.999285647\\
5	0.999863018\\
6	0.999438224\\
7	0.999326488\\
8	0.605352322\\
9	0.030724189\\
10	0.000668631\\
11	1.39e-05\\
12	2.91e-07\\
13	6.06e-09\\
14	1.23e-10\\
15	2.41e-12\\
16	4.57e-14\\
17	9.96e-16\\
18	4.08e-16\\
19	2.68e-16\\
20	2.28e-16\\
};
\addplot [color=mycolor1,solid,forget plot,line width=0.75pt]
  table[row sep=crcr]{2	0.999642986\\
3	0.99823563\\
4	0.999645073\\
5	0.998396249\\
6	0.999539274\\
7	0.996618714\\
8	0.045123364\\
9	0.000592841\\
10	6.88e-06\\
11	7.77e-08\\
12	8.18e-10\\
13	7.89e-12\\
14	6.92e-14\\
15	7.58e-16\\
16	3.15e-16\\
17	1.88e-16\\
18	1.11e-16\\
19	5.34e-17\\
20	1.44e-17\\
};
\addplot [color=mycolor1,dashdotted,line width=0.75pt]
  table[row sep=crcr]{2	0.999931729\\
3	0.999893266\\
4	0.999613936\\
5	0.999939385\\
6	0.999670964\\
7	0.999616076\\
8	0.999817125\\
9	0.999654067\\
10	0.999686428\\
11	0.999404806\\
12	0.999292285\\
13	0.998872801\\
14	0.998645849\\
15	0.997772243\\
16	0.999054526\\
17	0.993003349\\
18	0.386140996\\
19	0.036388711\\
20	0.001371847\\
};
\addlegendentry{$\sqrt{\epsilon_r}=3$};

\addplot [color=mycolor1,dashdotted,forget plot,line width=0.75pt]
  table[row sep=crcr]{2	0.999964218\\
3	0.999776297\\
4	0.999285647\\
5	0.999966452\\
6	0.999933411\\
7	0.999326488\\
8	0.999570943\\
9	0.999391602\\
10	0.998627032\\
11	0.999860755\\
12	0.999602272\\
13	0.999170908\\
14	0.617980638\\
15	0.028673085\\
16	0.000490949\\
17	8.82e-06\\
18	1.82e-07\\
19	3.93e-09\\
20	8.78e-11\\
};

\addplot [color=mycolor1,dashdotted,forget plot,line width=0.75pt]
  table[row sep=crcr]{2	0.999927112\\
3	0.999930104\\
4	0.999645073\\
5	0.996968257\\
6	0.998933972\\
7	0.99916459\\
8	0.99912089\\
9	0.999391322\\
10	0.999059787\\
11	0.999430341\\
12	0.999994031\\
13	0.119933308\\
14	0.001621911\\
15	1.73e-05\\
16	2.11e-07\\
17	2.69e-09\\
18	3.56e-11\\
19	4.59e-13\\
20	6.62e-15\\
};

%\draw[->,line width=0.25pt] (12.25,0.25) -- (17,0.6) {};
%\node[label={\scriptsize $0.6\cdot d_\text{Opt}$}] at (10.5,0.1) {};
%\draw[->,line width=0.25pt] (11.1,0.0325) -- (14.5,0.06) {};
%\node[label={\scriptsize $0.4\cdot d_\text{Opt}$}] at (9.35,0.015) {};
%\draw[->,line width=0.25pt,label={w}] (10.25,0.005) -- (13.5,0.007) {};
%\node[label={\scriptsize $0.2\cdot d_\text{Opt}$}] at (8.5,0.002) {};

\draw[<->,line width=0.25pt] (10.6,0.3) -- (17.25,0.6) node [pos=0.85,fill=none,xshift=0mm,yshift=0mm,below] {\scriptsize $0.6\cdot d_\text{Opt}$};
\draw[<->,line width=0.25pt] (9.25,0.0325) -- (14.5,0.06) node [pos=0.8,fill=none,xshift=0mm,yshift=0mm,below] {\scriptsize $0.4\cdot d_\text{Opt}$};
\draw[<->,line width=0.25pt] (8.75,0.004) -- (13.5,0.007) node [pos=0.7,fill=none,xshift=0mm,yshift=0mm,below] {\scriptsize $0.2\cdot d_\text{Opt}$};

\end{axis}
\end{tikzpicture}%
\label{fig:SqEpsilonr}\tikzexternaldisable}
\caption{Inverse condition number of LOS ULAs with dielectric medium, shape of the medium determined by a monotonic function, for different numbers of antennas $M=N$: \protect\subref{fig:OptFunctions}~Different functions, $0.8\cdot d_\text{Opt}$, $\sqrt{\epsilon_r}=2$; \protect\subref{fig:SqEpsilonr}~Function $z^2+\frac{1}{2}$ with small spacings and for two different $\sqrt{\epsilon_r}$ constants.}
\label{fig:ulaopt_mult}
\end{figure*}

\subsection{Numerical Results on other Arrangements}
For a higher number of antennas finding the optimal distances is even more difficult. We will thus pursue a slightly different approach. As in the previous section it will be assumed that $\matr{L}$ has Toeplitz structure. Then, we define the entries of the matrix to be given by
\begin{equation}
l_{mn} = l_\Delta \cdot |g\left(z\right)|
\end{equation}
where $l_\Delta$ is a constant and $|g\left(z\right)|$ is some function that describes the shape of the dielectric medium, where $z=|m-n|$.

With this setup we can then search over different $l_\Delta$ values and functions $g\left(z\right)$ in order to find $l_{mn}$ distances that achieve high inverse condition numbers $1/\kappa$. Fig.~\ref{fig:OptFunctions} shows the results for three different functions versus $N$, where we added the constraint $\max(l_{mn})-\min(l_{mn})\leq 2.5\lambda_0$. This constraint ensures that the $l_{mn}$ values are in a similar region which reduces implementation requirements. From the plot we can infer that $1/\kappa$ can be improved significantly over the standard free-space case depending on the function that is chosen. We chose monotonic functions with rapidly increasing slopes to achieve sufficient phase shifts under reasonable conditions. The quadratic function performs best as it is able to compensate approximately the path length that is lost due to the spacing reduction, given the right function parameters.

In Fig.~\ref{fig:SqEpsilonr} the behavior of the function $z^2+\frac{1}{2}$ is investigated further. We can see that the lower the desired spacing, the lower the number of antennas for which the optimal case is achieved. One remedy to that is to increase the value of $\sqrt{\epsilon_r}$, another would be lifting the constraint between minimal and maximal $l_{mn}$. This is intuitive since a reduction in spacing means that more phase shifting has to occur in the medium, because the phases of the free-space part become more similar.

In general, a steeper increase in distance values in $\matr{L}$ is necessary if a smaller spacing is desired. For example, we give the optimal vectors for $N=5$, $\sqrt{\epsilon_r}=2$ when $d=0.8\cdot d_\text{Opt}$ and $d=0.4\cdot d_\text{Opt}$ respectively, which were found to be
\begin{align}
\left(\matr{L}_{0.8}\right)_{1,n} &=
\lambda_0
\begin{bmatrix}
0.50 & 0.54 & 0.64 & 0.82 & 1.08 \\
\end{bmatrix}
\\
\left(\matr{L}_{0.4}\right)_{1,n} &=
\lambda_0
\begin{bmatrix}
0.50 & 0.58 & 0.84 & 1.26 & 1.84 \\
\end{bmatrix} \text{.} 
\end{align}
It needs to be pointed out that the determined paths in the medium do not need to be used on one side of the link, they can also be split and applied to transmitter and receiver side correspondingly. For some comments on practical issues consult the Appendix.

\section{Conclusion}
We proposed a deconstruction of the commonly used LOS MIMO channel matrix into a free-space propagation part and a phase shifting part. In general, the separate consideration of a phase shifting matrix can be used to improve the conditioning of the channel. We discussed how a dielectric medium, which is added to the propagation path, can potentially be used to achieve this phase shifting property. Based on different distances that the waves from different antennas travel in the medium, the LOS channel matrix is influenced differently. As such, the medium can be used to reduce the antenna spacing requirements for LOS MIMO systems, which can become prohibitively large in some cases. We showed that the optimal antenna spacing for ULAs, in terms of Shannon capacity, can be lowered compared with the case of just free-space propagation. The feasible array size reduction will be determined by the medium shapes and types that are practically realizable. The method can potentially be extended to the general case of two- or three-dimensional antenna arrangements, for which further improvements in array size are possible.

\appendix
\section*{Derivation of the Optimal Spacing with Rectangular Medium}

As stated in \emph{Eq.~\eqref{eq:innerproduct}} in the paper, the general criterion for orthogonality (and optimal channel matrix) is
\begin{align}
\left\langle \matr{h}_k,\matr{h}_l \right\rangle = \sum_{m=1}^{M-1} \exp\left(-j\frac{2\pi}{\lambda_0}(r_{mk}-r_{ml})\right) \cdot\exp\left(-j\frac{2\pi}{\lambda_0}(\sqrt{\epsilon_r}-1)(l_{mk}-l_{ml})\right)= 0 \text{.} \label{eq:innerproduct2}
\end{align}
By using
\begin{equation}
l_{mn} = t\cdot\frac{r_{mn}}{R} \label{eq:l_rec2}
\end{equation}
we get to
\begin{align}
\left\langle \matr{h}_k,\matr{h}_l \right\rangle &= \sum_{m=1}^{M-1} \exp\left(-j\frac{2\pi}{\lambda_0}\left(1+(\sqrt{\epsilon_r}-1)\frac{t}{R}\right)(r_{mk}-r_{ml})\right) \label{eq:ap1}\\
 &= \sum_{m=1}^{M-1} \exp\left(-j\frac{2\pi}{\lambda_0}\left(1+(\sqrt{\epsilon_r}-1)\frac{t}{R}\right)\cdot\frac{d_t d_r\cos\theta_t\cos\theta_r}{R}(k-l)m\right) \label{eq:ap2} \text{.}
\end{align}
When going from \eqref{eq:ap1} to \eqref{eq:ap2} one has to employ the steps mentioned in \cite{Bohagen2005a}. Earlier derivations of this without the tilting angle may be found in \cite{Gesbert2002,Haustein2003}. We need 
\begin{align}
r_{mn} =& \sqrt{ \left(R^2+(m-1) d_r\sin\theta_r-(n-1) d_t\sin\theta_t\right)^2 + \left((m-1) d_r\cos\theta_r-(n-1) d_t\cos\theta_t\right)^2 }\\
\approx& R + (m-1)d_r\sin\theta_r -(n-1)d_t\sin\theta_t + \frac{\left((m-1)d_r\cos\theta_r-(n-1)d_t\cos\theta_t\right)^2}{2R} \label{eq:approxr}
\end{align}
where compared with \cite{Bohagen2005a} we omit the azimuth angle, $\phi_r=0$. Further, $m-1$ and $n-1$ are used here to conform with the notation in our paper, i.e. $m=1...M$, $n=1...N$, which makes, however, no difference for the further calculations. When calculating $r_{mk}-r_{ml}$, assuming $M>N$, it should be visible that only the mixed term of the quadratic part in \eqref{eq:approxr} is useful, and that it is indeed the one seen in \eqref{eq:innerproduct2}. Then, following the same argument as \cite{Bohagen2005a} of a finite geometric series
\begin{align}
\left\langle \matr{h}_k,\matr{h}_l \right\rangle = \frac{\sin\left(\pi\left(1+(\sqrt{\epsilon_r-1})\frac{t}{R}\right)\frac{d_t d_r\cos\theta_t\theta_r}{\lambda_0 R}(l-k)M\right)}{\sin\left(\pi\left(1+(\sqrt{\epsilon_r-1})\frac{t}{R}\right)\frac{d_t d_r\cos\theta_t\theta_r}{\lambda_0 R}(l-k)\right)}\overset{!}{=} 0 \text{,}
\end{align}
for which one solution is
\begin{align}
\left(1+(\sqrt{\epsilon_r-1})\frac{t}{R}\right)\frac{d_t d_r\cos\theta_t\theta_r}{\lambda_0 R}M = 1 \text{,}
\end{align}
which leads in turn to the optimal antenna separation presented in \emph{Eq.~\eqref{eq:newspacing}} of the paper, after performing the same derivation for the case $N>M$.

\section*{Toeplitz Matrix Structure}

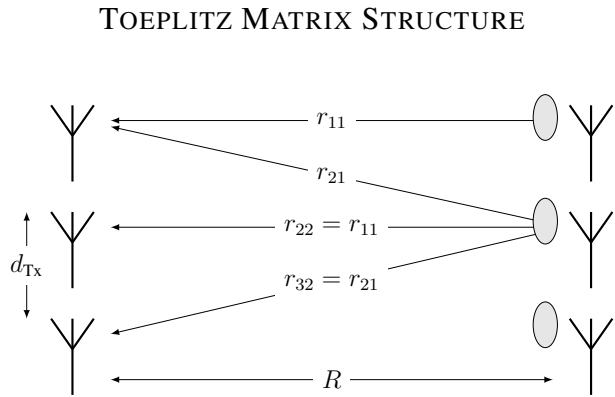
\begin{figure}[!ht]
\centering
\resizebox{0.5\textwidth}{!}{\begin{tikzpicture}[>=latex,font=\normalsize]

\usetikzlibrary{shapes}

\draw[line width=1pt] (1.5,0) -| (1.5,1.25);
\draw[line width=1pt] (1.15,1.25) -- (1.5,0.75);
\draw[line width=1pt] (1.85,1.25) -- (1.5,0.75);

\begin{scope}[shift={(0,1.75)}]

\draw[line width=1pt] (1.5,0) -| (1.5,1.25);
\draw[line width=1pt] (1.15,1.25) -- (1.5,0.75);
\draw[line width=1pt] (1.85,1.25) -- (1.5,0.75);

\end{scope}

\begin{scope}[shift={(0,3.5)}]

\draw[line width=1pt] (1.5,0) -| (1.5,1.25);
\draw[line width=1pt] (1.15,1.25) -- (1.5,0.75);
\draw[line width=1pt] (1.85,1.25) -- (1.5,0.75);

\end{scope}

\begin{scope}[shift={(8.5,0)}]

\draw[line width=1pt] (1.5,0) -| (1.5,1.25);
\draw[line width=1pt] (1.15,1.25) -- (1.5,0.75);
\draw[line width=1pt] (1.85,1.25) -- (1.5,0.75);

\begin{scope}[shift={(0,1.75)}]

\draw[line width=1pt] (1.5,0) -| (1.5,1.25);
\draw[line width=1pt] (1.15,1.25) -- (1.5,0.75);
\draw[line width=1pt] (1.85,1.25) -- (1.5,0.75);

\end{scope}

\begin{scope}[shift={(0,3.5)}]

\draw[line width=1pt] (1.5,0) -| (1.5,1.25);
\draw[line width=1pt] (1.15,1.25) -- (1.5,0.75);
\draw[line width=1pt] (1.85,1.25) -- (1.5,0.75);

\end{scope}

\end{scope}

\begin{scope}[shift={(-0.75,1.25)}]
\draw[<->,>=latex] (1.5,0) -- (1.5,1.75) node [pos=0.5, fill=white,xshift=0mm,yshift=0mm] {$d_\text{Tx}$};
\end{scope}

%\begin{scope}[shift={(9.25,1.25)}]
%\draw[<->,>=latex] (1.5,0) -- (1.5,3.5) node [pos=0.59, fill=white,xshift=0mm,yshift=0mm] {$D_\text{Rx}$};
%\end{scope}

\begin{scope}[shift={(0,3.5)}]
\draw[<->,>=latex] (2.125,1) -- (9.375,1) node [pos=0.5, fill=white,xshift=0mm,yshift=0mm] {$r_{11}$};
\end{scope}

\begin{scope}[shift={(0,1.75)}]
\draw[<->,>=latex] (2.125,1) -- (9.375,1) node [pos=0.5, fill=white,xshift=0mm,yshift=0mm] {$r_{22}=r_{11}$};
\end{scope}

\draw[<->,>=latex] (2.125,4.4) -- (9.375,2.8) node [pos=0.5, fill=white,xshift=0mm,yshift=0mm] {$r_{21}$};

\draw[<->,>=latex] (2.125,1) -- (9.375,2.7) node [pos=0.5, fill=white,xshift=0mm,yshift=0mm] {$r_{32}=r_{21}$};

\begin{scope}[shift={(0,-0.75)}]
\draw[<->,>=latex] (2.125,1) -- (9.375,1) node [pos=0.5, fill=white,xshift=0mm,yshift=0mm] {$R$};
\end{scope}

\node[ellipse, minimum width=0.2cm, minimum height=0.75cm,draw=black,fill=black!10] at (9.25,4.55) {};
\node[ellipse, minimum width=0.2cm, minimum height=0.75cm,draw=black,fill=black!10] at (9.25,2.85) {};
\node[ellipse, minimum width=0.2cm, minimum height=0.75cm,draw=black,fill=black!10] at (9.25,1.15) {};

%\node[] at (8.75,4.5) {\color{black}$l_{11}$};
%\node[] at (8.75,1.3) {\color{black}$l_{31}$};

%\node[align=center] at (8.75,5.25) {Medium, $\epsilon_r$};

%\begin{scope}[shift={(0,1.75)}]
%\draw[<->,>=latex] (8.5,1) -- (9.0,1) node [pos=0.5,above, fill=none,xshift=0mm,yshift=0mm] {$t$};
%\end{scope}

\end{tikzpicture}}
\caption{Symmetric ULA setup for Toeplitz structure visualization.}
\label{fig:setup2}
\end{figure}
Given the figure above it should be visible that in this case; meaning equidistant antenna spacing, no tilt between the arrays, symmetric medium with respect to the boresight of the antenna and same medium shape for each antenna; the distance matrix in free-space $\matr{R}$ as well as in the medium $\matr{L}$ can be represented as a Toeplitz structure
\begin{equation}
\matr{R}=
\begin{bmatrix}
r_{11} & r_{21} & r_{31} & \cdots & r_{M1} \\
r_{21} & r_{11} & r_{21} & \cdots & r_{(M-1)1} \\
r_{31} & r_{21} & r_{11} & \cdots & r_{(M-2)1} \\
\vdots & \vdots & \ddots & \ddots & \vdots \\
r_{M1} & r_{(M-1)1} & r_{(M-2)1} & \cdots & r_{11} \\
\end{bmatrix} \text{ and }
\matr{L}=
\begin{bmatrix}
l_{11} & l_{12} & l_{13} & \cdots & l_{1N} \\
l_{12} & l_{11} & l_{12} & \cdots & l_{1(N-1)} \\
l_{13} & l_{12} & l_{11} & \cdots & l_{1(N-2)} \\
\vdots & \vdots & \ddots & \ddots & \vdots \\
l_{1N} & l_{1(N-1)} & l_{1(N-2)} & \cdots & l_{11} \\
\end{bmatrix} \text{.}
\label{eq:toeplitz2}
\end{equation}
For $\matr{L}$, each row represents the paths that each of the transmit antennas sees to one particular receive antenna. The first row mirrored at $l_{11}$ would also represent the total shape of the medium per antenna. As stated in the paper, setups with different numbers of antennas on each side can also be investigated with this structure. For example, consider removing the third antenna on the receiver side. We simply need to remove the last row in $\matr{L}$ and could then perform the optimization. It should also be visible here, that it is not necessary to have a dielectric in front of every antenna, i.e., certain rows would be zero vectors, which can be helpful in certain situations.

\section*{Practicality}

\subsection{LOS MIMO}
We will provide a small general discussion of the LOS MIMO scheme here, to clarify why the pure LOS (non-scattering) assumption is practically relevant. LOS MIMO is based on the fact that, even for long range links, the propagating waves should be modeled as spherical rather than plane waves, theoretical aspects of that fact have been discussed in \cite{Bøhagen2009}. Considering that fact, the channel matrix is entirely determined by the phase relations that occur between antennas due to tiny differences in the propagation time. Initially, the spatial multiplexing achieving case for that channel was deemed most useful for short-range communications due to the fact that, as link range increases (with a fixed carrier frequency) the antenna spacing also has to increase, see \emph{Eq.~\eqref{eq:optULA}} in the paper. Furthermore, as link range increases the impact of scatters may also increase, even for LOS setups when considering antennas with low directivity.

With the advent of millimeter-wave communication this situation has changed a bit. First, the higher carrier frequencies allow for a reduction in required spacing, or, equivalently for longer link ranges (we generally consider few hundreds of meters). Furthermore, due to the required high antenna gains, fewer scatters will occur in general additionally justifying the LOS MIMO assumption. Finally, various measurements have shown that even if reflections or scattering occurs, the reflected power is usually largely reduced. An implementation of the concept for a \SI{41}{\metre} link at \SI{60}{\giga\hertz} can be found in \cite{Sheldon2008a}. A discussion of the applicability for satellite communication is given in \cite{Knopp2008}.

Note that beamforming is also a very useful technique for these LOS links, but we are here and in the present paper mostly concerned with the pure spatial multiplexing case as has also been investigated in the previous works.

\subsection{Dielectric Medium}
For a practical implementation further aspects, other than the optimal distances in the medium, have to be investigated. We have not considered effects such as reflection and refraction, which will influence the performance if the thickness of the medium is large with respect to the wavelength. In light of that it is also interesting to investigate how materials with very high dielectric constants
%, $\sqrt{\epsilon_r}=100$ materials exist (capacitor dielectrics), 
behave since they relax the design constraints significantly. Additionally, one could investigate how a material with different dielectric constants may be beneficial for this approach, see \cite{Imbert2015}. The fundamental limit of this technique will be set by physical constraints, e.g., possible medium shapes and their accuracy, see \cite{Al-Joumayly2011} for another lens implementation.

Finally, it is of interest how the determined optimal length differences in the dielectric material could be achieved, when considering the real antenna aperture. One easy way to imagine this is to think of effective antenna apertures. The dielectric material with the specified thickness could be placed at such a distance in front of the receiving antenna that the effective apertures of the transmitting antennas do not overlap.

%\section{}

%\section*{Acknowledgment}

\ifCLASSOPTIONcaptionsoff
  \newpage
\fi



\bibliographystyle{IEEEtran}
\bibliography{references}

\end{document}